# Perils and opportunities of AI for public climate services

Francesca Larosa[1,2], Sandro Calmanti[3], Matteo De Felice[4], Marcello Petitta[5]

1 Euro-Mediterranean Center on Climate Change (CMCC), Venice, Italy

2 Ca' Foscari University, Venice, Italy

3 ENEA, Rome, Italy

4 Independent contribution

5 RomaTre University, Rome, Italy

*Corresponding author: francesca.larosa@cmcc.it

## Abstract

The rapid integration of artificial intelligence (AI), particularly generative and machine-learning systems, in climate services development and provision is transforming how climate information is produced, accessed, and interpreted. AI-enabled climate services offer significant opportunities for process automation, optimisation, tailoring, and the translation of complex climate data into actionable knowledge for diverse users. However, the growing reliance on privately controlled algorithms, data infrastructures, and computing facilities raises critical concerns related to governance, transparency, equity and economic sustainability, among others. While the technical architecture of climate services increasingly relies on advanced machine learning and large-scale climate models, their legitimacy as public services may be weakened in the absence of arrangements that prioritise public value and long-lasting societal benefit over profit maximisation. The tension between the public mandate of climate services and the private nature of AI infrastructure challenges traditional notions of openness. Using a consensus-based elicitation method coupled with the PESTLE framework, we analyse the political, economic, social, technological, legal, and environmental risks and opportunities shaping the co-production value chain of AI-enabled climate services. We draw policy-relevant conclusions for data, model and infrastructure supporting the next generation of upstream and downstream climate services as critical applications for society. Our work presents potential and plausible challenges and benefits for a pan-European, decentralised public climate service built on sovereign AI infrastructure and open governance principles.

## 1. Introduction

From data collection to processing and interpretation, artificial intelligence (AI) is transforming the monitoring and assessment of environmental change[1]. Traditional approaches to data collection and climate information provision struggle to support decisions with the adequate level of precision, granularity and tailoring[1]. Climate impacts are growing in frequency and magnitude virtually anywhere on Earth[2–4], boosting demand for accessible, usable and bespoke approaches to anticipate and prepare for physical risks, while also planning for the future. Generative and agentic AI promise to retrieve, process and deliver context-relevant and user-oriented climate insights at scale[5]. While traditionally populated by national meteorological services who use public infrastructure, the availability of cloud-based advanced computational tools has opened opportunities to private actors offering climate services (CS). Climate services involve the provision of climate information to assist decision-making[6]. By supporting planning, monitoring, assessment and revisions of decisions where physical and transition climate risk may alter outcomes and processes[4]. CS work across different temporal and spatial scales shifting the risk management logic from reactive to proactive. Their explosive development has motivated the establishment of multi-level governance frameworks (such as the Global Framework for Climate Services, GFCS), incentives to stimulate their upscaling[7] and the persistent efforts to standardise and define them[8]. With the advent of machine learning first, and AI then, a growing number of private companies were born to support critical decisions by combining proprietary and public information. The market for CS is hard to quantify due to the heterogeneity of activities, products and skills involved.

Estimates for Europe (EU28) in 2016/2017 reported a total revenue stream around 7 billion EUR for climate services and 6.1 billion EUR for weather services[9]. These seem misaligned to 2024 private estimates where services accounted for 5.8 billion USD globally[10]. The AI component accelerates and expands the market per user base, type of activities and diversity of products, but it also creates new opportunities and opens to potential risks.

This paper explores, identifies and discusses the risks associated with the growing dependency of public services, such as CS, on critical infrastructures, skills and capabilities that are owned and managed privately. We apply a PESTLE analytical framework and assign qualitative scores (from 1 to 4) to identify risks and opportunities of variable urgency. Our work achieves three goals. First, it reframes climate services as strategic tools under current EU legislation. Second, it exposes overlooked techno-economic s with significant societal and security implications. Third, it proposes and suggests risk-mitigation strategies to build EU digital sovereignty while simultaneously strengthening competitiveness through climate resilience. Our work outlines a policy roadmap that leverages AI to strengthen climate aware decision-making, positioning climate resilience as a central policy objective rather than a mere co-benefit of innovation and economic security.

## 1.1 Climate services as contested governance territories

CS are commonly defined as the generation, provision, and contextualisation of climate information and knowledge to assist decision-making at multiple levels of society, spanning individual, organisational, and governmental scales[11]. This framing establishes CS as boundary activities linking climate science to actionable choices in agriculture, water management, energy, health, and disaster risk reduction[12]. Because the same information can simultaneously function as a tradable knowledge product and as a determinant of collective safety (for example, by informing flood and heatwave preparedness), its appropriate governance regime remains contested. Three positions dominate the debate:

**The profit-driven position.** CS have been theorised as Knowledge-Intensive Business Services (KIBS): activities whose value derives from specialised expertise and whose diffusion follows market logics of openness or proprietary restriction depending on firm strategy[13]. KIBS are economically valuable activities providing non-financial services to both private and public actors[14]. They are highly human-capital and knowledge-intensive[15], mainly intangible and difficult to standardise[13]. Empirical work on the European climate service market applies business-model canvases to climate service providers, treating information as a value proposition to be captured through licensing, tailoring, or subscription[16]. Under this logic, access is a competitive variable rather than an entitlement. **The public-purpose position.** A contrasting view holds that climate information underpins societal preparedness and is therefore inseparable from national and local security. The GFCS explicitly links climate services to disaster risk reduction, food security, and health resilience as public missions[17]. National meteorological and hydrological services have historically operated as public infrastructure funded to protect life and property and to support treaty obligations such as aviation and maritime safety. This position treats restriction of access as a societal risk in itself, favouring open, state-guaranteed provision[18]. **The economist position.** A third view argues the market should self-regulate, but recognises that the climate service sector is atypically supply-driven: providers, largely public research bodies, generate products before users articulate demand, producing a persistent "usability gap"[19,20]. From this standpoint, intervention should focus not on ownership but on demand-building (engagement, co-production, and literacy), so that price signals and competition can eventually allocate resources efficiently[16,21].

The CS landscape is shaped by the (co-)existence of two highly differentiated modes of operations. Upstream providers, the Copernicus Climate Change Service, national meteorological and hydrological services, and regional environmental protection agencies, generate, quality-control and disseminate foundational climate information under public mandates. They respond to a largely public-purpose call. Downstream providers, often termed purveyors or intermediaries[22,23], tailor this information into sector-specific products for end users: physical risk assessments for financial institutions, agronomic advisories, adaptation planning support for municipalities. The tension between the upstream and downstream segments can be illustrated by the insurance sector, where large companies with substantial financial and technological resources may blur the upstream–downstream boundary by developing proprietary climate risk information, potentially exacerbating information asymmetries and contributing to adverse selection and moral hazard[24]. Empirical mappings of the European market confirm that this segmentation is not merely analytical: of the 371 climate service providers identified across EU Member States, public institutions dominate the upstream while private actors concentrate in downstream tailoring activities[25,26], a division of labour that the Copernicus programme explicitly anticipates in its value-chain design.

These contested territories are not necessarily mutually exclusive but describe different degrees of vertical integration of the same value chain. The upstream services (observation networks, reanalysis, and baseline projections) exhibit classic public-good characteristics (i.e., non-rivalry, high fixed cost, security relevance) and are empirically dominated by public institutions even where private markets exist downstream[25,26]. Upstream institutions control the full production stack: ECMWF, for instance, operates proprietary high-performance computing and, through the Common Cloud Infrastructure at the Bologna Data Centre, dedicated GPU capacity for machine-learning workloads, all governed by intergovernmental agreements. When such institutions adopt AI methods, they do so on infrastructure they own, under quality-assurance regimes they define.

The downstream layer (e.g., sector-specific advisories and decision products) can sustain competitive KIBS-style provision without compromising public safety, provided the underlying data and infrastructure remains open[27]. Downstream services rely on externally procured components: open data from the Climate Data Store, foundation models accessed through commercial APIs, and hosting on hyperscale cloud platforms. The leading private climate-analytics firms operate on AWS, Microsoft Azure or Google Cloud Platform, a dependency structure consistent with the broader European pattern, in which a small number of US providers serve the large majority of the cloud market and underpin critical state functions[28,29]. As CS remain largely "supply push" rather than genuinely co-produced[30,31] pure market self-regulation underperforms without public investment in demand articulation, while purely proprietary models risk excluding vulnerable users from safety-critical information. A layered governance model based on public critical infrastructure for core data and early-warning functions and competitive markets for tailored value-added services, therefore reconciles efficiency with the security imperative underlying the public-purpose argument.

A clear mismatch exists between mandate and means: climate services are expected to function as public, safety-critical infrastructure, yet the AI capabilities on which their next generation depends (foundation models, GPU compute and cloud hosting) are increasingly owned, priced and governed by a small number of private, largely non-European providers. Because upstream institutions still control their production stack, this dependency materialises chiefly downstream, where no public authority owns, audits or can guarantee over time the infrastructure through which climate information reaches its users.

Three bodies of scholarship address portions of this mismatch, but none captures its intersection. The CS literature has framed the principal barrier between data and decision as epistemic, a “usability gap” to be closed through co-production and user engagement[32,33], implicitly assuming that once information is usable, access to the means of producing it is unproblematic. The emerging literature on AI in weather and climate prediction has concentrated on model skill and, more recently, on ethics, bias and trustworthiness[34,35], with growing attention to the migration of forecasting capability from public to private actors[36], but little analysis of the market structure through which that capability reaches end users. Finally, the digital sovereignty literature has documented the strategic dependencies created by hyperscaler concentration and the limits of European counterinitiatives such as Gaia-X[28,37], yet treats the public sector generically, without attending to sectors, such as climate services, whose outputs qualify as critical infrastructure under the CER Directive[38]. The result is a lack of clarity of where intervention is needed: policy attention has concentrated on upstream data openness, which is largely secured, while the downstream segment, where proprietary models, extraterritorial data regimes (e.g.,US CLOUD Act) and unilateral terms-of-service revisions intersect with decisions affecting public safety, remains unaddressed by both the EU AI Act, under which climate services are ambiguously categorised, and the climate services governance literature itself.

Climate services are critical public infrastructure: essential physical and cyber tools and services whose limitation or destruction would lead to debilitated or hampered societal outcomes. They are such by mandate: they support resilience and protection of people and assets across different temporal and spatial scales. Climate services also proactively transform business practices: they embed climate risks within operations and planning with significant consequences along the supply chain within and between sectors. In a world of increasing capital investments from large technology companies, with 2026 capital expenditure expected to surpass by 75% the 400 billion USD recorded in 2025[15], governance of the computing infrastructure enabling critical public applications becomes urgent from a cyber, physical and societal security standpoint.

### 1.2 Artificial intelligence in climate services

AI has a profound impact on climate services: it improves climate forecasts and predictions[39], processing heterogeneous and complex data sources[40] and facilitating knowledge sharing in quasi real-time[41]. Fast advancements in computational infrastructure, as well as in model design and physics-informed machine learning models, dramatically decrease the time and costs of running large-scale simulations, promising to democratising climate information at scale. The use of LLMs has the power to increase resilience[42,43], improve data accessibility[44], explore existing knowledge[45] and support adaptation planning. Chatbots are mediating interfaces: by collecting, synthetising, summarising and elaborating knowledge from diverse information sources[46], they make climate services and adaptation literacy more accessible even to remote and vulnerable communities[47]. However, being inherently socio-technical, AI systems generate and accelerate new and existing issues, which shall not remain overlooked. The literature has mainly focused on three core issues.

First, the integration of AI in operational workflows which assist and improve decision-making remains a challenge. A gap between usefulness and usability persists. Scholars have identified trustworthiness[39], explainability[48] and legitimacy[49] of AI-generated content as key drivers of this usability gap. Furthermore, ML and AI systems are cheaper to prototype, but more expensive to run[50], widening the "valley of death", a stage after a concept has been shown to work in a controlled setting and before it is reliably used, funded and maintained in real decisions[51]. The usability gap is also highly influenced by the institutional readiness in receiving and deploying AI-based systems in established routines[52]. Digital and climate services literacy often proceed at different speeds and this divide increases for complex services value chains. In adaptation, the valley of death is also prompted by market dynamics. The funding landscape for adaptation services is fragmented and often public-dominated[53] in which open data policies act as pre-conditions to services to exist[54]. However, in some contexts and sectors, commercialisation itself represent a limitation factor to mass adoption and public services are advocated as ideal[55].

Second, the expansion of AI-based climate services perpetuates and expands inequalities[56] both per geography[57] and group[58]. Inputs (i.e., data, agendas and priorities) are largely designed and developed in the Global North with scarce and still too limited inclusion practices of Global South needs[57]. Climate services are the combination of technological, scientific and economic specialisation, often concentrated in limited areas of the world[59]. AI accelerates these clusters with significant effects on outputs: preparedness, resilience and adaptation in mounting climate changes.

Third, the over-reliance on AI-generated insights leads to a false perception of co-production. AI systems operate opaquely and influence behaviour in conscious and nonconscious ways. Taking finance as an example, statistical machine learning algorithms excel at prediction tasks[60,61], whereas genAI tools and chatbots deliver advice and recommendations[62] intended to reduce decision uncertainty[63]. In climate services, we have not yet assessed how significant these expanded sensory, cognitive-perceptive and behavioural-performative capabilities may be at altering, changing and nudging human behaviour. However, we know that machines have the capacity to mitigate controversial positions[64], shape belief formation[65] and produce more relatable content which influence judgement[66].

The existence of these trade-offs has different magnitude at upstream and downstream climate services level. The upstream segment is constrained by and benefits from punctual rulebooks and procedures which are hard to change. The downstream segment, especially on the private side, is strongly encouraged, but standards are advocated to ensure the quality of both the AI and climate sciences components.

## 2. Methods: charting a complex forward-looking environment

Both upstream and downstream climate services aim at delivering actionable climate information by transforming and translating climate data into knowledge which the users find useful[21]. Actionability is an abstract term in the field of climate intelligence[67] and AI can improve and expand it by integrating in three layers: data, model and infrastructure of the service. AI can collect and homogenise diverse datasets, filling existing gaps or contributing to enriching the evidence base[5]. On the model layer, AI methods support detection, prediction and impact assessment, but struggle with understanding the causal relationships between physical variables[39]. On the infrastructure side, AI scales computation, dramatically decreasing the cost of simulations and projections without sacrificing accuracy[68]. It also enables new forms of communications and results delivery by integrating genAI and web-hosted services into fully operational applications. We focus on these three layers, data, model and infrastructure, and we apply the PESTLE framework as a strategic tool to analyse the macro-environmental exogenous factors which influence the operation of a business or a service[69].

The PESTLE framework is useful to systematically categorise risks and opportunities of the fast-changing AI-enabled climate services landscape. It serves as an evaluation tool of (presumed or confirmed) political (P), economic (E), social (S), technology (T), legal (L) and environmental (E) factors which reinforce or hamper public climate services. The political dimension focuses on ownership, autonomy and control of AI-enabled climate intelligence. The climate services literature often frames these aspects in terms of governance of climate information[70], but mainly targets the data and model aspects of it[71]. The economic dimension addresses the intertemporal economic sustainability of the service. Climate services literature has explicitly developed around valuation of benefits from the use of these tools[72–74]. However, both empirical and theoretical studies often neglect an equity and distributional component of climate services deployment across time and space. The social dimension involves the framing and assessment of climate services as social practices. The literature has explored the determinants of demand for climate services: rather than assuming their adoption as solely linked to technical aspects, climate services are largely shaped by social and institutional conditions[75]. The "next generation" of climate services shall recognise how intertwined climate challenges are to other social issues[76], which must be carefully identified. The technological component looks at the technical and material requirements for climate services to work as cutting-edged innovations. These include innovative data collection practices (e.g., Earth observations) and data analysis suits[77]. Both can benefit from large-scale AI integration[39], provided transparency and control is guaranteed. The legal dimension involves the standardised practices and rules to leapfrog climate services benefits and utilisation. Standardisation of climate services is a fairly new area of work[78], but the growing integration of AI methods and practices calls for stronger emphasis on ethics, privacy and transparency requirements. Finally, the environmental dimension informs how climate services contribute to the conservation, protection, preservation and risk mitigation of the environmental capital. While the literature explicitly addresses climate adaptation[79], the ecological implications of their use is often framed as part of the ecosystem services and nature-based solutions literature. This siloed approach hampers the assessment of climate services benefits and over-emphasise product-based rather than process-based approaches[71].

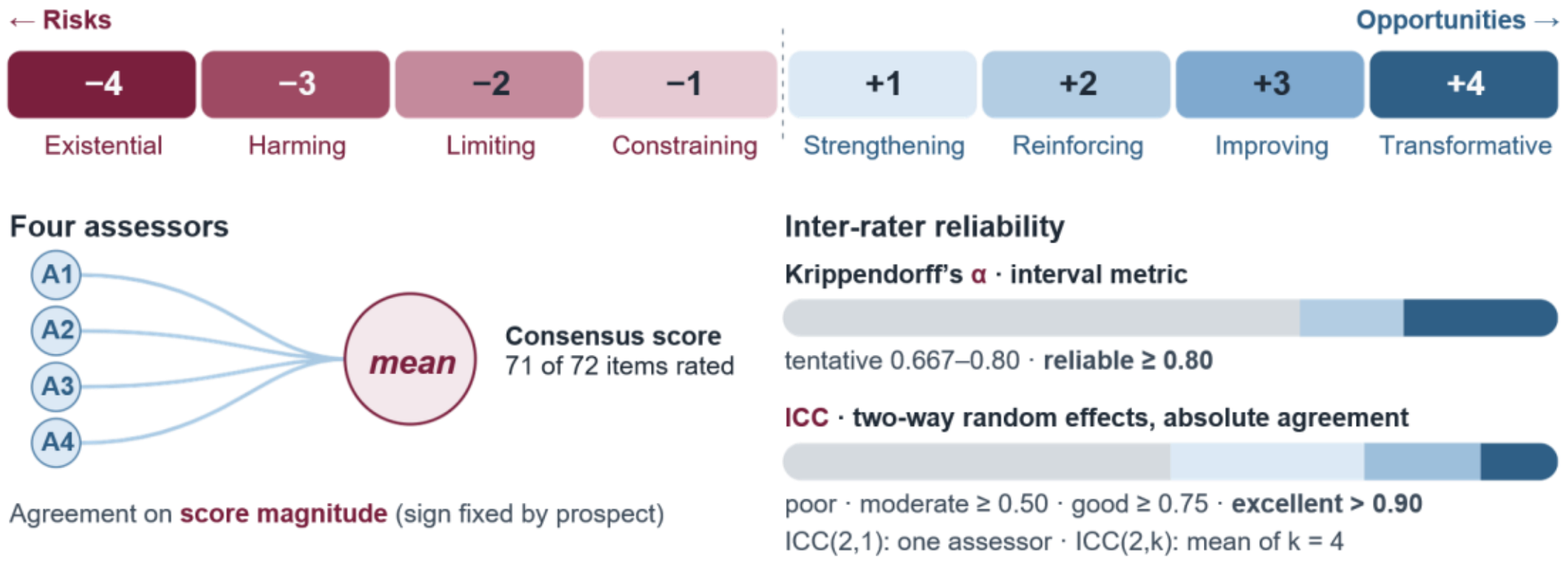

**Figure 1.** Scores and process to assess the severity of risks and opportunities.

We measure the severity of the risk or opportunity on a scale from 1 to 4 (Figure 1) with negative values for risks and positive values for opportunities (Table 1) for 72 items (3 domains × 6 PESTLE factors × 2 prospects × 2 segments) as four independent assessors. The consensus score for each item is the mean of the available ratings (71 as one item was left unassessed). We computed Krippendorff's $\alpha$ with an interval metric[80] and the intraclass correlation coefficient (ICC). The ICC expresses the proportion of total score variance attributable to true differences between the items being rated, as opposed to variance arising from the assessors[81]. It ranges from 0 (i.e., scores carry no information about item differences) to 1 (i.e., assessors are interchangeable). We report ICC(2,1), the expected reliability of a single assessor's score, and ICC(2,k), the reliability of the mean of the k = 4 assessors, which is the quantity that underlies the consensus scores. Krippendorff's α expresses agreement as one minus the ratio of observed to chance-expected disagreement among the scores assigned to each item. A value of 1 indicates perfect agreement and 0 indicates agreement no better than chance. Unlike the ICC, α accommodates missing ratings and can be defined for nominal, ordinal or interval data, and we applied it with an interval metric[82]. Following Krippendorff[82], $\alpha \geq 0.80$ is conventionally taken as reliable and $0.667 \leq \alpha < 0.80$ as acceptable only for tentative conclusions, while ICC values are interpreted as poor (<0.50), moderate (0.50–0.75), good (0.75–0.90) and excellent (>0.90)[82]. Both coefficients were computed on score magnitude, because the sign of each score is fixed by the prospect (challenge or opportunity) and would artificially inflate agreement. Confidence intervals were obtained by bootstrapping the items (2,000 resamples). This assessment was conceived as a pilot exercise involving four independent assessors representing different perspectives on climate services. Although the small number of assessors limits the generalisability of the results, the diversity of their perspectives provides an opportunity to explore how risks and opportunities are perceived and prioritised differently. In this context, inter-assessor disagreement is not merely a limitation of the scoring process, but also a potentially informative outcome, helping to identify the most contested aspects of AI integration into climate services and priorities for further investigation with a broader community of stakeholders.

Table 1. Scoring table for both risks and opportunities

| Score | Risk | Opportunity |
|---|---|---|
| 1 | **Constraining:** Imposes a constraint on the use of climate services without affecting core functionalities; repair is possible at low cost. Creates friction or sustainability/administrative constraints, but does not significantly reduce service availability or reliability. | **Strengthening:** Consolidates existing climate services at low cost, with relatively modest return on investment. Stabilises or legitimises the current public-service baseline. |
| 2 | **Limiting:** Limits the scope, usability or availability of climate services; repair requires higher costs. Reduces who can use the service, for what purpose, or with what confidence. | **Reinforcing:** Enhances currently available climate services, for example through efficiency gains, but with limited qualitative improvements. Makes existing services cheaper, cleaner, faster or easier to operate. |
| 3 | **Harming:** Potentially damages currently available functionalities; recovery is possible but costly and long-term. Can undermine trust, correctness, preparedness, equity or continuity of existing services. | **Improving:** Creates measurable incremental improvement in availability, usability or accuracy relative to existing climate services.Produces concrete improvements in service performance, access, quality or decision relevance. |
| 4 | **Existential/Irreversible**: Produces an unrecoverable failure of climate services. Would permanently destroy essential public climate-service capacity. | **Transformative:** Enables a new generation of climate services, different from what is available today. Changes the service paradigm, typically through public AI models, sovereign platforms, or new user-facing capabilities. |

# 3. Results

## 1. Charting risks and opportunities for AI in public climate services

As AI integration in the climate services production chain increases[83], risks and opportunities simultaneously manifest themselves, highlighting that challenges are not just technology-related, but largely governance linked. We use the PESTLE framework to unpack the key challenges across data, model and infrastructure components of AI in climate services. Our goal is to provide a non-exhaustive, but representative list of both risks and opportunities per PESTLE segment to suggest whether policy interventions can support the existing mandate and vision for a democratic, open and public climate services ecosystem. Reactive policies responding to fast technological progress may not be adequate to ensure climate services respond to public mandate and collective needs. Instead, the analysis of potential threats and the assessment of their direction, helps the creation of policy guardrails for flexible governance mechanisms. The PESTLE dimensions are not operating independently from one another, but synergically. First, we elaborate on each of them. Second, we use a consensus-based approach (i.e., the Delphi method) to discuss their impact on public climate services development and deployment discriminating between upstream and downstream levels when relevant.

**Political (P) mandate, strategic autonomy and democratic oversight.** At European level, upstream climate services enabled by Earth observation infrastructures (and particularly Copernicus) have been established as strategic political priorities[84]. Their mandate explicitly frames climate data as a public good, committed to open and free access[85]. However, as political priorities change in times of geopolitical tensions, security and control, climate data become instrumental for AI-driven systems linked to competitiveness and critical infrastructure resilience[43]. This expanding scope may gradually constrain openness through selective access, creating pressures on the original mandate. Similarly, AI model development is shaped by concerns over strategic autonomy and dependency on non-European foundation models. This is highly relevant for downstream climate services built on foreign generative AI systems due to privacy and cybersecurity concerns. The EU AI Act provides a regulatory basis for risk-based oversight, documentation requirements, and accountability mechanisms that help embed public-interest objectives into model governance. However, the ecosystem of self-assessment audits which *de facto* the EU puts in place cannot guarantee adequate democratic oversight[86]. Furthermore, the growing dependence of climate services on privately owned hyperscale digital infrastructures may shift strategic decisions on AI development towards market competition and profitability, rather than collective societal needs and long-term public value. For instance, commercial incentives may favour the development of high-resolution AI-based climate risk assessment tools for profitable sectors, such as finance and insurance, over equally important capabilities supporting climate adaptation in vulnerable communities with limited purchasing power[87]. European high-performance computing initiatives represent an opportunity to strengthen sovereign and publicly controlled computing capacity, reinforcing climate services as essential public goods.

**Economic (E) sustainability and economies of scale**. Climate data infrastructures require long-term and secure funding for collection, curation, storage, and maintenance[31]. At downstream level, the case for sustained public budgets depends on socio-economic benefit evidence whose methodology is not standardised yet[88]. At the upstream level, Europe's consolidated practice in Earth observation demonstrates the feasibility and value of treating climate data as shared infrastructure[85]. At the same time, advanced AI model development demands substantial, recurring investments in data center clusters, model training and innovation support to meet competitiveness goals[84]. Without coordinated public investments, reliance on proprietary tools has an impact on switching costs, thereby increasing vendor lock-in risks and generating significant power imbalances[89]. Treating AI models as shared public infrastructure can reduce duplication, enable economies of scale and strengthen long-term infrastructure planning. Moreover, it can trigger positive cascade effects by stimulating a downstream market through the deployment of upstream and quality assured systems. Liability and risk allocation pose additional economic challenges when AI-supported climate services inform high-stakes decisions (i.e., critical infrastructure investment, disaster management and early warning): there remain unclear accountability issues along the value chain, which may impact on public budgets[90,91]. Importantly, the development of shared and common, but EU sovereign AI systems become a powerful pre-condition for the protection against misalignment in the age of agentic AI[90], reducing against malicious threats beyond climate.

**Social (S) acceptance through trust and inclusion**. Although climate data are non-personal, AI-enabled services often combine environmental datasets with socio-economic or citizen-generated data, posing privacy and ethical complexities. At downstream level, participatory practices such as citizen science and co-produced datasets support legitimacy and trust of AI tools, which in turn are essential elements to promote AI adoption and scaling[92]. At downstream level, citizen science can enrich the knowledge base for bespoke climate services[93] and become the tool to raise awareness about adaptation needs, planning and risk management, while also reducing biases in data collection[94]. Participatory and bottom-up data gathering methods are also useful to integrate indigenous and traditional knowledge[95] preserving the identity of specific groups and connecting climate services to other agendas, such as the Sustainable Development Goals (SDGs)[96]. However, challenges remain, especially at upstream level, where quality assurance and database consistency are key to avoid cascade and critical errors[97]. In the data domain, participatory approaches must guarantee representativeness of the population, focusing especially on data scarce areas or groups[94]. At the same time, citizen science should not be used in isolation, but integrated to enhance credibility and saliency of official data sources[94]. This is particularly useful to avoid model biases once training is completed, which may influence risk rankings and eligibility for adaptation measures. We know from the climate services literature that co-development of services is a strong determinant of adoption. When AI systems are involved, the co-development approach can be expanded from the data and model design phases. At downstream level, this would build familiarity and promote ease of use, while also supporting transparency.

**Technological (T) progress to support standardised transparent systems.** Transparency in data collection, cleaning, processing and sharing is a key component of climate services innovation. As climate services are deployed in often critical contexts in which extensive human judgement is required[58], standardised approaches to metadata and data processing opens new opportunities for responsible adoption[58]. The use of standards is not limited to the data aspects only, but it involves the four components of climate services: decision context, ecosystem of actors and co-production process, delivery mode and evaluation and knowledge systems[78]. Technology is an enabler of the standardisation pathways[98]: common standards and AI-ready data architectures strengthen upstream reusability and make it easier for downstream services to combine climate information with sector-specific datasets without losing traceability. Explainable AI can support interpretability, ensuring critical scrutiny of model biases. In climate services, these challenges must be carefully considered as black box innovation erodes value in the long-run[99] and reduces credibility and salience[58]. AI is also instrumental to power and monitor quality assurance protocols across different upstream and downstream climate services[100]. When generative AI is used as a source of climate data, continuous validation is required to avoid hallucinations[101]. Generative AI is also the engine behind chatbot-enabled climate services, where misinterpretation risks are high, especially when English is not the main language[42]. At downstream level, these challenges can lead to maladaptation practices and inaccurate strategies[42]. To avoid them, the use of multi-model AI ensembles can increase robustness of outputs[100], but the integration of human scrutiny and regular output evaluation are needed.

**Legal (L) framework to ensure accountability and improve governance.** The legal architecture of AI in Europe is firmly rooted in the EU AI Act (2024[102]) and it unfolds in the Apply AI strategy[103]. This legal framework is informed by the established regulations in ICT, digitalisation and twin transition (e.g., the Chips Act, the GDPR and the Data Act, among others)[104]. From a public perspective, the rise of private AI systems and infrastructure poses questions about sovereignty, competition and control[105]. The reliance on third-party data sources, for example, introduces legal uncertainty related to licensing, intellectual property rights, data reuse conditions, and cross-border data governance, particularly in an international climate data ecosystem[106]. AI training rights and downstream derivative products are emerging legal grey zones in which regulation must balance public interest and security issues. In the model layer, unclear legal ownership (IPR) and responsibilities related to errors or misuse, complicate accountability in AI-enabled climate services, particularly when models are developed collaboratively. Platform and interface ownership determines who controls the user relationship, presentation of uncertainty, terms of use, and redress pathways, altering the co-production processes and potentially harming strategic decisions[89]. However, opportunities exist: climate services (and climate intelligence as a whole) paves the road for a comprehensive public-oriented regulatory framework using standards as key leverages. These shall cover product- and process-based innovations carefully balancing trade-offs.

**Environmental (E) alignment, digital footprints and shadow prices.** The energy and material demands of data storage, processing, and AI model training can threaten climate goals[107], but their applications can accelerate mitigation and adaptation efforts[108]. These environmental trade-offs are not addressed by shared and agreed principles or frameworks between the green and digital policy landscapes. However, the expansion of AI capabilities and the race to equip countries with frontier models is widening the so-called “environmental debt”: resource consumption today may not justify the benefits in the future[109]. Aligning AI development, net-zero policies and resilience goals in Europe has become a political priority under mounting grassroot protests. Publicly optimised upstream data platforms can reduce duplicate downloads and processing, allowing downstream users to create value from data-proximate workflows rather than repeatedly moving large datasets[110]. This public-led leapfrogging[111] can optimise for energy efficiency, low-carbon operation aligning them with broader sustainability objectives. Moreover, planning workflows can match reduced carbon profiles, lowering the pressure on the electricity grid as well[112]. At application level, AI systems which support climate and environmental goals shall be incentivised. These include foundation models for climate sciences, as well as applications to foster adaptation and mitigation planning[108].

## 2. Assessing severity and urgency of possible changes

Overall (Table 2), we reach moderate agreement with $\alpha = 0.25$ (bootstrap 95% CI 0.16-0.33) and single-rater ICC(2,1) = 0.26, rising to ICC(2,k)=0.58 for the mean of the four raters. Individual ratings are thus not interchangeable, showing differences in value judgement among assessors. However, the consensus mean is moderately reliable, providing the seeds for a more extensive and in-depth analysis. We assessors converge more on challenges ($\alpha$=0.29) than on opportunities ($\alpha$=0.18), but opportunities (mean +2.56) were rated more intensely than challenges (-2.25). We assigned the same score on only 4 to 72 items (6%): the spread was at most one point on 30 items (42%) and two or more points on 42 (58%), reaching three points on two items. Pairwise, assessors fall within one scale point on 72-90% of items (exact agreement 27-52%), so disagreement is mostly of degree rather than direction.

Table 2. Agreement statistics

| Metric | Value |
|---|---|
| Krippendorff's α, signed/magnitude | 0.90/0.25 (95% CI 0.16-0.33) |
| A, challenges/opportunities | 0.29/0.18 |
| ICC(2,1) / ICC(2,k) | 0.26/0.58 |
| Items with identical scores / spread ≤ 1 and ≥2 | 4 (6%) / 30 (42%) / 42 (58%) |
| Pairwise mean absolute difference; exact; within 1 point | 0.59–1.03; 27–52%; 72–90% |
| Mean item standard deviation | 0.75 |

Differences (Figure 2) between the three domains are small with infrastructure/QA scored highest (-2.46/+2.71), followed by model (-2.22/+2.69) and data (-2.08/+2.29). Summing mean opportunity and challenge scores per domain and segment, model upstream has the most favourable balance (+3.42), whilst data has the least (+1.25 in both segments). From our assessment, the data case rests on consolidating existing public infrastructure (many scores of 1-2) rather than transforming it. This is in line with the historical climate services development, which heavily focused on data systems and architecture to support decision-making. The PESTLE factors differ more (Table 3). Political is the most intense on both sides (-3.00/+3.13), followed by Technological (-2.33/+2.96); Environmental. Social and Legal items are consistently milder.

Being a nascent and infant landscape, in which AI is in constant evolution, technological opportunities dominate over threats. This is particularly visible in the downstream market, where competitive markets, open models and shared infrastructure can spur innovation in adaptation and resilience with bespoke services. However, the governance of the technologies underpinning public climate services emerges as the most significant challenge. **Upstream**, the absence of strong political commitment can constrain the development of sovereign AI capabilities for climate services, encompassing control over both infrastructure and processes, including quality assurance, while limiting the potential to realise economies of scale. Similarly, equity considerations (S) cannot be adequately addressed without coordinated upstream interventions capable of shaping market conditions and transmitting appropriate downstream incentives. **Downstream**, challenges are generally less pronounced and more evenly distributed across dimensions, with notable exceptions in the political and economic domains. The political dimension is particularly ambivalent: concerns over control, sovereignty and institutional mandates underpin both the most significant risks (i.e., security-driven mission drift in data provision and private ownership of critical infrastructure) and some of the most substantial opportunities, notably the development of democratically governed European AI models. In this context, common standards for AI deployment could facilitate harmonised practices for data collection, maintenance and sharing, while strengthening quality assurance across the climate services value chain.

The **economic dimension** is the only one in which downstream challenges outweigh opportunities (net score of −0.75, corresponding to −0.25 in each domain). This imbalance is primarily driven by dependence on proprietary model architectures (Model, −3.25) and unresolved liability for AI-generated errors (Infrastructure/QA, −3.00). The current downstream model development landscape remains fragmented, with competitive dynamics potentially encouraging costly duplication of efforts and investment patterns that yield limited returns to the real economy. Reliance on proprietary APIs and closed model architectures risks creating persistent technological dependencies for downstream applications, increasing switching costs, constraining interoperability and weakening the bargaining power of public institutions. Conversely, economic opportunities are more pronounced upstream, where the development of a publicly governed, EU-wide foundation model could generate **economies of scale and positive spillovers across the climate services value chain**. By providing a shared and sovereign technological foundation, such an investment could reduce duplication, lower barriers to downstream innovation and enable the development of diverse applications without reinforcing dependence on proprietary infrastructures. This would allow upstream public investment to generate cascading economic and societal benefits while preserving the scope for competitive differentiation among downstream providers.

Table 3. Mean scores per factor

| Factor | Risk | Opportunity | Spread ≥2 |
|---|---|---|---|
| Political | -3.00 (0.83) | +3.13 (1.01) | 10 |
| Technological | -2.33 (0.78) | +2.96 (0.65) | 7 |
| Economic | -2.46 (0.58) | +2.58 (0.64) | 4 |
| Legal | -1.88 (0.69) | +2.42 (0.75) | 6 |
| Social | -1.97 (0.75) | +2.17 (1.02) | 9 |
| Environmental | -1.88 (0.58) | +2.13 (0.68) | 6 |

The upstream and downstream segments exhibit heterogeneous risk and opportunity profiles, with differences varying across PESTLE dimensions. Overall, downstream scores are marginally more pronounced for both challenges (−2.33 versus −2.17 upstream) and opportunities (+2.63 versus +2.50 upstream), suggesting that risks tend to intensify as climate services move closer to end users. In particular, inaccurate, incomplete or misused data can introduce biases whose severity increases downstream (Data–Technological, −2.00 upstream versus −3.00 downstream). These risks are further compounded by opaque AI interfaces, which may conceal underlying errors and undermine the transparency and reproducibility of results. Similarly, biases embedded in AI models can propagate into adaptation planning and investment decisions, with potentially significant economic and financial consequences, including maladaptation and the dissemination of misleading information.

Conversely, certain quality-assurance dimensions are more pronounced upstream, particularly technological opportunities associated with Infrastructure/QA (+3.25 upstream versus +2.25 downstream) and social challenges (−2.75 versus −2.00). These differences underscore the strategic importance of upstream governance in establishing the conditions for reliable and equitable downstream service provision. Sovereign public infrastructure can safeguard the continuity of foundational data and model services while ensuring that downstream access, prioritisation and service levels remain subject to democratic accountability. Moreover, transparent quality-assurance frameworks and public risk-pooling mechanisms could have substantial economic implications by clarifying the allocation of liability across the value chain, reducing duplicated assurance costs and supporting more reliable downstream decisions. Realising these benefits requires continuous upstream monitoring of authoritative databases, complemented by systematic expert review and iterative validation to ensure the accuracy, consistency and reliability of the information underpinning downstream applications.

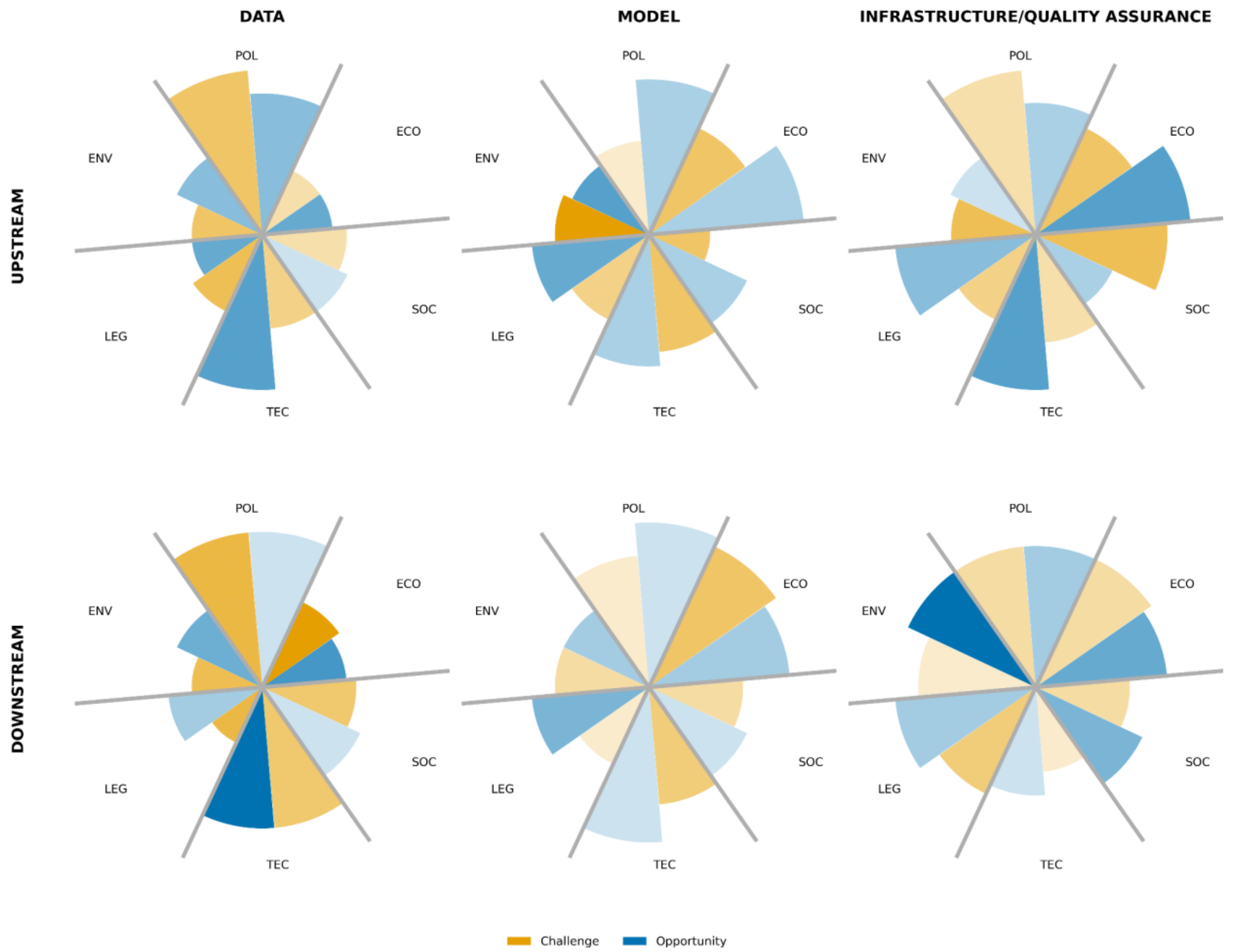

**Figure 2. Consensus-based assessment of risks and opportunities.** Wedge length reflects the mean and, as we read it, saturation reflects consensus (inversely related to the SD). The most saturated wedges (SD = 0) are Data–Technological opportunity downstream, Data–Economic challenge downstream, Model–Environmental challenge upstream and Infrastructure/QA–Environmental opportunity downstream. The palest wedges are mostly Political and Social opportunities.

## 4. Discussion and conclusions

Despite being explorative and qualitative, our analysis sheds light on hidden and non-obvious loopholes which can hamper a public climate service for the future, limiting adaptation and shrinking public-led innovation. Disagreement between us assessors reveals some evolving dynamics, which are currently unfolding and require careful scrutiny. These disagreements are also the baselines for discussion along four patterns of change.

The transformative power of institutional mandate is our first point of disagreement per magnitude of effect on AI-driven public services. Public climate services require a strong political vision, which unleash coherent investments and capacity building programmes. Europe's consolidated habit of pooling together resources through Copernicus programmes and Destination Earth (among others) can become the backbone of those critical climate services infrastructure enabled by AI, but they cannot be standalone. Existing efforts such as OpenEuroLLM[113] must engage in a dialogue with the climate community from an early stage, fulfilling a purposefully twin transition. Furthermore, the institutional mandate without strong governance structures does not create the most favourable conditions for a public-led flourishing private sector. The second point of disagreement relates to the framework required to build truly inclusive and equitable climate services. On the data side, we identify bottom-up data collection methods (through citizen science) to ensure representativeness of the population and extended analyses of vulnerabilities. However, we noted disagreement in assessing the usability of such data points and note that the quality assurance protocols to ensure data is deployed effectively must be regularly discussed, updated and revised. On the model side, AI training must ensure biases are mitigated and removed: their presence can affect priority interventions and adaptation strategies. Computing infrastructure and facilities shall support the whole continent: the effective use of climate services affects public budgets and enhances existing competitive advantages. The integration of AI into the climate services value chain can widen existing inequality gaps, dividing the continent rather than unifying it. Third, the fragile equilibrium between political mandate and control shall not be overstated. Strategic autonomy, security and sovereignty should not constraint model transparency. Private and public actors can work new collaborative, rather than competitive, business models to ensure that IPR is protected but not at the expense of public interest. Finally, we left a legal caveat open with respect to liability in the face of inaccurate or wrong decisions. When infrastructure, interfaces, or middleware are owned or operated by private actors, responsibility for service reliability, accuracy of presentation, user guidance, and failure management becomes fragmented. This blurs accountability for errors, outages, or misleading representations of climate information, complicating the enforcement of public-law obligations and weakening users' ability to identify who is legally responsible for the service.

Our analysis is far from being exhaustive and its breadth could be enhanced by involving a larger pool of experts/assessors. However, it offers a framework and a platform to discuss challenges in designing, developing and deploying public critical infrastructure such as climate services, in the age of AI. As noted, both upstream and downstream market segments face and benefit from political, economic, social, technological, legal and environmental aspects. The use of a consensus-based approach to discuss them and chart their potential impacts support the creation of informed risk mitigation strategies.

References

1. Kuznetsov, I. *et al.* Transforming climate services with LLMs and multi-source data integration. *npj Clim. Action* **4**, 97 (2025).
2. Bellavite Pellegrini, C., Pellegrini, L. & Vismara, S. Climate change and systemic risk: Evidence from financial intermediaries. *Technol. Forecast. Soc. Change* **222**, 124408 (2026).
3. Chancel, L., Mohren, C., Bothe, P. & Semieniuk, G. Climate change and the global distribution of wealth. *Nat. Clim. Chang.* **15**, 364–374 (2025).
4. Lesk, C. *et al.* Compound heat and moisture extreme impacts on global crop yields under climate change. *Nat. Rev. Earth Environ.* **3**, 872–889 (2022).
5. Koldunov, N. & Jung, T. Local climate services for all, courtesy of large language models. *Commun. Earth Environ.* **5**, 13 (2024).
6. Doblas-Reyes, F. J. *et al.* Standardisation of equitable climate services by supporting a community of practice. *Clim. Serv.* **36**, 100520 (2024).
7. Reveco Umaña, C. *et al.* The challenge of upscaling climate services: Towards a common framework. *Clim. Serv.* **42**, 100662 (2026).
8. Street, R. B. Towards a leading role on climate services in Europe: A research and innovation roadmap. *Clim. Serv.* **1**, 2–5 (2016).
9. Howard, S., Howard, S. & Howard, S. Quantitative market analysis of the European Climate Services sector – The application of the kMatrix big data market analytical tool to provide robust market intelligence. *Climate Services* vol. 17 100108 at https://doi.org/10.1016/j.cliser.2019.100108 (2020).
10. Strategic revenue insights. Environmental Intelligence Service Market Size, Future Growth and Forecast 2033. *Strategic revenue insights* https://www.strategicrevenueinsights.com/industry/environmental-intelligence-service-market (2013).
11. Vaughan, C. & Dessai, S. Climate services for society: Origins, institutional arrangements, and design elements for an evaluation framework. *Wiley Interdiscip. Rev. Clim. Chang.* **5**, 587–603 (2014).
12. Hewitt, C. D., Stone, R. C. & Tait, A. B. Improving the use of climate information in decision-making. *Nat. Clim. Chang.* **7**, 614–616 (2017).
13. Miles, I. *et al. Knowledge-Intensive Business Services: Users, Carriers and Sources of Innovation.* https://pure.manchester.ac.uk/ws/portalfiles/portal/32800224/FULL_TEXT.PDF (1995).
14. Strambach, S. Innovation Processes and the Role of Knowledge-Intensive Business Services (KIBS). in *Innovation Networks: Concepts, Challenges in the European Perspective* (eds. Koschatzky, K., Kulicke, M. & Zenker, A.) (Physica-Verlag HD, Berlin, 2001). doi:10.1007/978-3-642-57610-2.
15. Brenner, T., Capasso, M., Duschl, M., Frenken, K. & Treibich, T. Causal relations between knowledge-intensive business services and regional employment growth. *Reg. Stud.* **52**, 172–183 (2018).
16. Larosa, F. & Mysiak, J. Business models for climate services: An analysis. *Clim. Serv.* 100111 (2019) doi:10.1016/j.cliser.2019.100111.
17. Hewitt, C. D. *et al.* Making Society Climate Resilient: International Progress under the Global

Framework for Climate Services. *Bull. Am. Meteorol. Soc.* **101**, E237–E252 (2020).
18. Petitta, M. *et al.* From global data to local action: A participatory framework for national wind, solar, and hydropower atlases based on observations, climate projections and artificial intelligence. *Clim. Serv.* **42**, 100663 (2026).
19. Lourenço, T. C., Swart, R., Goosen, H. & Street, R. The rise of demand-driven climate services. *Nat. Clim. Chang.* **6**, 13–14 (2015).
20. Vaughan, C., Buja, L., Kruczkiewicz, A. & Goddard, L. Identifying research priorities to advance climate services. *Clim. Serv.* **4**, 65–74 (2016).
21. Bessembinder, J. *et al.* Need for a common typology of climate services. *Clim. Serv.* **16**, 100135 (2019).
22. Brasseur, G. P. & Gallardo, L. Climate services: Lessons learned and future prospects. *Earth's Futur.* **4**, 79–89 (2016).
23. Guentchev, G., Palin, E. J., Lowe, J. A. & Harrison, M. Upscaling of climate services – What is it? A literature review. *Clim. Serv.* **30**, 100352 (2023).
24. Boomhower, J., Fowlie, M., Gellman, J. & Plantinga, A. *How Are Insurance Markets Adapting to Climate Change? Risk Classification and Pricing in the Market for Homeowners Insurance*. https://www.nber.org/papers/w32625 (2024) doi:10.3386/w32625.
25. Cortekar, J., Themessl, M. & Lamich, K. Systematic analysis of EU-based climate service providers. *Clim. Serv.* **17**, 100125 (2020).
26. Tart, S., Groth, M. & Seipold, P. Market demand for climate services: An assessment of users' needs. *Clim. Serv.* **17**, 100109 (2020).
27. Larosa, F. & Perrels, A. *ASSESSMENT OF THE EXISTING RESOURCING AND QUALITY ASSURANCE OF CURRENT CLIMATE SERVICES Grant Agreement 730500 EU-MACS European Market for Climate Services Deliverable 1.2*. http://eu-macs.eu/wp-content/uploads/2017/07/EUMACS_D12_EXISTING-RESOURCING-AND-QUALITY-ASSURANCE-OF-CURRENT-CLIMATE-SERVICES.pdf (2017).
28. Blancato, F. G. The cloud sovereignty nexus: How the European Union seeks to reverse strategic dependencies in its digital ecosystem. *Policy and Internet* **16**, 12–32 (2024).
29. Blancato, F. G. & Carr, M. The trust deficit. {EU} bargaining for access and control over cloud infrastructures. *J. Eur. Public Policy* **33**, 936–967 (2026).
30. Weichselgartner, J. & Arheimer, B. Evolving Climate Services into Knowledge–Action Systems. *Weather. Clim. Soc.* **11**, 385–399 (2019).
31. Bruno Soares, M. & Buontempo, C. Challenges to the sustainability of climate services in Europe. (2019).
32. Alexander, M. & Dessai, S. What can climate services learn from the broader services literature? *Clim. Change* 1–17 (2019) doi:10.1007/s10584-019-02388-8.
33. Jacob, D. *et al.* Co-production of climate services: challenges and enablers. *Front. Clim.* **7**, 1507759 (2025).
34. McGovern, A., Ebert-Uphoff, I., Gagne, D. J. & Bostrom, A. Why we need to focus on developing ethical, responsible, and trustworthy artificial intelligence approaches for environmental science. *Environ. Data Sci.* **1**, e6 (2022).
35. Ben-Bouallégue, Z. & others. The rise of data-driven weather forecasting: A first statistical assessment of machine learning-based weather forecasts in an operational-like context. *Bull. Am. Meteorol. Soc.* **105**, E864--E883 (2024).
36. Martin, A. & Lin, C. Regimes of Scale in {AI} Meteorology. at https://doi.org/10.48550/arXiv.2604.06000 (2026).
37. Baur, A. European ambitions captured by American clouds: digital sovereignty through {Gaia-X}? *Information, Commun. \& Soc.* **29**, 417–434 (2026).
38. European Parliament and Council of the European Union. Directive (EU) 2022/2557 of the European Parliament and of the Council of 14 December 2022 on the resilience of critical entities and repealing Council Directive 2008/114/EC. *Official Journal of the European Union* vol. L 333 164–198 at https://eur-lex.europa.eu/eli/dir/2022/2557/oj (2022).
39. Camps-Valls, G. *et al.* Artificial intelligence for modeling and understanding extreme weather and climate events. *Nat. Commun. 2025 161* **16**, 1919- (2025).
40. Cheong, S.-M., Sankaran, K. & Bastani, H. Artificial intelligence for climate change adaptation. *Wiley Interdiscip. Rev. Data Min. Knowl. Discov.* **12**, (2022).
41. Acemoglu, D., Lin, T., Ozdaglar, A. E. & Siderius, J. How AI Aggregation Affects Knowledge. *SSRN Electron. J.* (2026) doi:10.2139/ssrn.6527034.
42. Rutan, R. *et al.* An AI visual chatbot system for climate heat risk communication and decision-making in at-risk agricultural communities. *Clim. Serv.* **44**, 100736 (2026).
43. Reichstein, M. *et al.* Early warning of complex climate risk with integrated artificial

intelligence. *Nat. Commun.* **16**, 2564 (2025).
44. DestinE. Build Destination Earth's weather and climate chatbot with us! | Destination Earth. *Destination Earth Newspage* https://destine.ecmwf.int/news/build-destination-earths-weather-and-climate-chatbot-with-us/ (2024).
45. ClimateADAPT. Ask Climate-ADAPT | Discover the key services, thematic features and tools of Climate-ADAPT Climate-ADAPT. https://climate-adapt.eea.europa.eu/en/ai-search-assistant (2026).
46. Wickberg, A. & Gärdebo, J. Computation, data and AI in Anthropocene history. *Hist. Technol.* **39**, 328–346 (2023).
47. Larosa, F. & Wickberg, A. Artificial Intelligence can help Loss and Damage only if it is inclusive and accessible. *npj Clim. Action* **3**, 59 (2024).
48. Sahoh, B. & Choksuriwong, A. The role of explainable Artificial Intelligence in high-stakes decision-making systems: a systematic review. *J. Ambient Intell. Humaniz. Comput.* **14**, 7827–7843 (2023).
49. Haibe-Kains, B. *et al.* Transparency and reproducibility in artificial intelligence. *Nature* **586**, E14–E16 (2020).
50. Sculley, D. *et al.* Hidden Technical Debt in Machine Learning Systems. in *Advances in Neural Information Processing Systems* vol. 28 (Curran Associates, Inc., 2015).
51. National Research Council. *From Research to Operations in Weather Satellites and Numerical Weather Prediction: Crossing the Valley of Death*. (The National Academies Press, Washington, DC, 2000). doi:10.17226/9948.
52. Chen, M. & Wang, C. Crossing the valley of death: A fault and coupling verification framework for technology-business integration in dual-node incubation systems. *J. Innov. Knowl.* **18**, 101072 (2026).
53. Rubio-Martin A., M. C. M. P.-V. M. G.-P. A. C. L. L. F. M. J. Structuring Climate Service Co-Creation Using a Business Model Approach. at https://doi.org/https://doi.org/10.1029/2021EF002181 (2021).
54. European Commission. Focused business models and open-data policies are key to accelerating uptake of climate services. at https://cordis.europa.eu/article/id/422475-focused-business-models-and-open-data-policies-key-to-accelerating-uptake-of-climate-services (2020).
55. Petitta, M., Calmanti, S. & De Felice, M. Climate Services as Public Infrastructure: Lessons from European Research Partnerships. in *EGU General Assembly 2026* (Vienna, Austria, 2026). doi:10.5194/egusphere-egu26-6991.
56. Mozaffari, A. *et al.* The rise of AI in weather and climate information and its impact on global inequality. *npj Clim. Action* (2026) doi:10.1038/s44168-026-00412-z.
57. Meng, M., Hamm, N. A. S., Yang, L. E., Wang, J. & Chan, F. K. S. Global disparities and structural imbalances in AI integration for urban climate resilience research. *npj Clim. Action* (2026) doi:10.1038/s44168-026-00431-w.
58. Williams, H. T. P. *et al.* Developing trustworthy AI for climate services at scale. *npj Clim. Action* **5**, 74 (2026).
59. Larosa, F. & Mysiak, J. Mapping the landscape of climate services. *Environ. Res. Lett.* (2019) doi:10.1088/1748-9326/ab304d.
60. KELLY, B., MALAMUD, S. & ZHOU, K. The Virtue of Complexity in Return Prediction. *J. Finance* **79**, 459–503 (2024).
61. for Development Africa, A. I. Artificial Intelligence for Climate Action: Advancing Mitigation and Adaptation in Developing Countries. *UNFCCC Technol. Exec. Comm.* **Techical p**, (2024).
62. Chandra, M. *et al.* Longitudinal Study on Social and Emotional Use of AI Conversational Agent. *arXiv Comput. Sci.* **1**, (2025).
63. Mbanyele, W. Generative AI and Investment Efficiency: Early Evidence Using Textual Analysis. *SSRN Electron. J.* (2025) doi:http://dx.doi.org/10.2139/ssrn.5081571.
64. Costello, T. H., Pennycook, G. & Rand, D. G. Durably reducing conspiracy beliefs through dialogues with AI. *Science (80-. ).* **385**, (2024).
65. Matz, S. C. *et al.* The potential of generative AI for personalized persuasion at scale. *Sci. Rep.* **14**, 1–16 (2024).
66. Glickman, M. & Sharot, T. How human–AI feedback loops alter human perceptual, emotional and social judgements. *Nat. Hum. Behav.* **9**, 345–359 (2025).
67. Jagannathan, K., Buddhavarapu, S., Ullrich, P. A. & Jones, A. D. Typologies of actionable climate information and its use. *Glob. Environ. Chang.* **82**, 102732 (2023).
68. Lam, R. *et al.* Learning skillful medium-range global weather forecasting. *Science (80-. ).* **382**, 1416–1421 (2023).
69. Vilas, H. & Jain, B. S. P. PESTLE ANALYSIS. *Int. J. Adv. Res.* **13**, 608–612 (2025).

70. Golding, N. *et al.* Developing national frameworks for climate services: Experiences, challenges and learnings from across Europe. *Clim. Serv.* **37**, 100530 (2025).
71. Findlater, K., Webber, S., Kandlikar, M. & Donner, S. Climate services promise better decisions but mainly focus on better data. *Nat. Clim. Chang. 2021 119* **11**, 731–737 (2021).
72. Crochemore, L. *et al.* A Framework for Joint Verification and Evaluation of Seasonal Climate Services across Socioeconomic Sectors. *Bull. Am. Meteorol. Soc.* **105**, E1218–E1236 (2024).
73. An-Vo, D. A., Radanielson, A. M., Mushtaq, S., Reardon-Smith, K. & Hewitt, C. A framework for assessing the value of seasonal climate forecasting in key agricultural decisions. *Clim. Serv.* **22**, 100234 (2021).
74. Tesfaye, A., Hansen, J., Kassie, G. T., Radeny, M. & Solomon, D. Estimating the economic value of climate services for strengthening resilience of smallholder farmers to climate risks in Ethiopia: A choice experiment approach. *Ecol. Econ.* **162**, 157–168 (2019).
75. Jesse, S. & Mathias, V. What are climate services for? Social practices, adaptation, and the dynamics of demand. *Clim. Serv.* **42**, 100640 (2026).
76. Jacobs, K. L. & Street, R. B. The next generation of climate services. *Clim. Serv.* **20**, 100199 (2020).
77. Lamptey, B. *et al.* Challenges and ways forward for sustainable weather and climate services in Africa. *Nat. Commun. 2024 151* **15**, 2664- (2024).
78. Doblas-Reyes, F. J. *et al.* Standardisation of equitable climate services by supporting a community of practice. *Clim. Serv.* **36**, 100520 (2024).
79. Cheong, S. M., Sankaran, K. & Bastani, H. Artificial intelligence for climate change adaptation. *Wiley Interdiscip. Rev. Data Min. Knowl. Discov.* **12**, e1459 (2022).
80. Krippendorff, K. *Content Analysis : An Introduction to Its Methodology*. (2004).
81. Shrout, P. E. & Fleiss, J. L. Intraclass correlations: uses in assessing rater reliability. *Psychol. Bull.* **86**, 420–428 (1979).
82. Hayes, A. F. & Krippendorff, K. Answering the Call for a Standard Reliability Measure for Coding Data. *Commun. Methods Meas.* **1**, 77–89 (2007).
83. World Meteorological Organization. Key Takeaways from the Learning Session: AI for Forecasting, Climate Services and Early Warning. Malawi's Experience. at https://community.wmo.int/media/news/key-takeaways-from-learning-session-ai-forecasting-climate-services-and-early-warning-malawis (2026).
84. von der Leyen, U. 2026 State of the Union Address. at https://ec.europa.eu/commission/presscorner/detail/en/SPEECH_26_1868 (2026).
85. Copernicus Programme. Setting the Course Ahead: Copernicus Releases the 2028--2034 Strategy for Its In-Situ Component. at https://insitu.copernicus.eu/news-and-events/news-folder/setting-the-course-ahead-copernicus-releases-the-2028-2034-strategy-for-its-in-situ-component (2026).
86. Hartmann, D., de Pereira, J. R. L., Streitbörger, C. & Berendt, B. Addressing the regulatory gap: moving towards an EU AI audit ecosystem beyond the AI Act by including civil society. *AI Ethics 2024 54* **5**, 3617–3638 (2024).
87. OECD. *Climate Adaptation Investment Framework*. (OECD Publishing, Paris, 2024). doi:10.1787/8686fc27-en.
88. Delpiazzo, E. *et al.* Co-evaluation of climate services. A case study for hydropower generation. *Clim. Serv.* **28**, 100335 (2022).
89. Rikap, C. & Weko, S. A green transition orchestrated from Big Tech clouds? *Globalizations* **7731**, 1–20 (2025).
90. Lagarde, C. Where AI risks meet. at https://www.ecb.europa.eu/press/key/date/2026/html/ecb.sp261001~cf3c630379.en.html (2026).
91. Dou, W. W., Goldstein, I. & Ji, Y. *AI-Powered Trading, Algorithmic Collusion, and Price Efficiency*. https://www.nber.org/papers/w34054 (2025) doi:10.3386/w34054.
92. Afroogh, S., Akbari, A., Malone, E., Kargar, M. & Alambeigi, H. Trust in AI: progress, challenges, and future directions. *Humanit. Soc. Sci. Commun. 2024 111* **11**, 1568- (2024).
93. Neset, T. S. *et al.* Co-designing a citizen science climate service. *Clim. Serv.* **24**, 100273 (2021).
94. Fraisl, D. *et al.* Why citizen science is now essential for official statistics. *Commun. Sustain. 2026 11* **1**, 2- (2026).
95. ALBATROSS Project. Citizen Science Toolkit. at https://albatross-project.eu/project/citizen-science-toolkit/.
96. Borghys, K., Vandercruysse, L., Veeckman, C., Temmerman, L. & Heyman, R. Localizing the sustainable development goals in smart and sustainable cities: how can citizen-generated data support the local monitoring of SDGs? A case study of the Brussels Capital Region. *Front.*

*Environ. Sci.* **12**, 1369001 (2024).
97. Grandin, M. *et al.* Citizen Science in Space and Atmospheric Sciences: Opportunities and Challenges. *Surv. Geophys. 2025* 1–35 (2025) doi:10.1007/S10712-025-09888-6.
98. Hewitt, C. D. *et al.* Recommendations for Future Research Priorities for Climate Modeling and Climate Services. *Bull. Am. Meteorol. Soc.* **102**, E578–E588 (2021).
99. Delpiazzo, E. *et al. Technical Briefs and Compendium on Assessment of Values and Benefits*. https://zenodo.org/records/22204008 (2026) doi:10.5281/zenodo.22204008.
100. Bratic, G., Gobron, N. & Cappucci, F. *AI-Driven Review of Quality Information within Copernicus Climate Change and Atmosphere Monitoring Services: Gaps and Recommendations*. https://publications.jrc.ec.europa.eu/repository/handle/JRC147495 (2026) doi:10.2760/2341032.
101. Malekela, A. A. *et al.* Generative artificial intelligence for climate information services. *Discov. Appl. Sci. 2026 86* **8**, 684- (2026).
102. EUR-LEX. *The EU AI Act*. (2024).
103. European Commission. *Apply AI Strategy: Communication from the Commission to the European Parliament and the Council*. https://eur-lex.europa.eu/legal-content/EN/TXT/?uri=CELEX:52025DC0723 (2025).
104. Ebert, K., Alder, N., Herbrich, R. & Hacker, P. AI, Climate, and Regulation: From Data Centers to the AI Act. *ArXiv* **abs/2410.0**, null (2024).
105. Rikap, C. 'Digital Public Infrastructure': Cultural hegemony and states' technological dependence. *Big Data Soc.* **13**, (2026).
106. Novelli, C., Casolari, F., Hacker, P., Spedicato, G. & Floridi, L. Generative AI in EU law: Liability, privacy, intellectual property, and cybersecurity. *Comput. Law Secur. Rev.* **55**, 106066 (2024).
107. Gaffney, O. *et al.* The Earth alignment principle for artificial intelligence. *Nat. Sustain.* **8**, 467–469 (2025).
108. Larosa, F. *et al.* Halting generative AI advancements may slow down progress in climate research. *Nat. Clim. Chang.* **13**, 497–499 (2023).
109. Turliuk, J. & Sterman, J. *The Net Climate Impact of Artificial Intelligence (AI): Balancing Current Costs with Future Climate Benefits*. https://ssrn.com/abstract=6689098 (2026) doi:https://dx.doi.org/10.2139/ssrn.6689098.
110. Abernathey, R. P. *et al.* Cloud-Native Repositories for Big Scientific Data. *Comput. Sci. Eng.* **23**, 26–35 (2021).
111. Watson, J. & Sauter, R. Sustainable innovation through leapfrogging: A review of the evidence. *Int. J. Technol. Glob.* **5**, 170–189 (2011).
112. Chen, X., Wang, X., Colacelli, A., Lee, M. & Xie, L. Electricity Demand and Grid Impacts of AI Data Centers: Challenges and Prospects. at https://doi.org/10.48550/arXiv.2509.07218 (2025).
113. OpenEuroLLM Consortium. OpenEuroLLM: Transparent AI for Europe. at https://openeurollm.eu/ (2026).